\documentclass[journal]{IEEEtran}
\ifCLASSINFOpdf
  \usepackage[pdftex]{graphicx}
  \graphicspath{{../pdf/}{../jpeg/}}
  \DeclareGraphicsExtensions{.pdf,.jpeg,.png}
\else
\fi

\usepackage{amsmath}

\usepackage{subfig}
\usepackage{amssymb}

\usepackage{tikz}
\usetikzlibrary{shapes,arrows}

\begin{document}

\title{Use of adaptive filtering techniques and deconvolution to obtain low range sidelobe samples}

\author{Mohit Kumar,~\IEEEmembership{Student Member,~IEEE,}
       V Chandrasekar,~\IEEEmembership{Fellow,~IEEE,}}

\maketitle

\begin{abstract}
In this paper the use of adaptive filtering techniques to obtain better peak sidelobe suppression and integrated sidelobe energy will be discussed with regard to weather radars and obtaining better sensitivity with this technique. The performance of these new coefficient sets obtained with adaptive filter (using RLS optimization) will be discussed and presented. They will also be compared with the existing techniques and their peak sidelobe levels.
\end{abstract}


\section{Introduction}
\IEEEPARstart{I}{n} this paper,we introduce a novel sidelobe suppression technique with adaptive filter optimization. It is capable of mitigating the unwanted sidelobe power of the pulse compression processing and equally applicable to observations of both distributed targets as in case of weather and point targets as well. With additive white gaussian noise, for a point target, a matched filter is known to have best possible signal to noise ratio (SNR) gain, but its higher sidelobe levels can mask the weak nearby targets or atleast influence them to a level for incorrect parameter identification (power, velocity etc). With the help of windowing such as kaiser, a good compromise can be obtained between mainlobe broadening, which reduces resolution, and sidelobe reduction. But even using this, sidelobe suppression below -35 to -40dBc (below the carrier peak power) is very difficult to obtain. This might lead to masking of weak echoes in vicinity by the higher power echoes. We would use the Least Squares (LS) based approach along with mismatched filters which is optimal in mean square error (MSE) sense to obtain better sidelobe characteristics. It was shown in \cite{Bharadwaj2012} that the peak sidelobe levels can be effectively reduced using longer mismatched filters, so we try to use a conjunction of adaptive filter and mismatched filter techniques.
\subsection{Signal Model}
Let \textbf{S} be the transmit convolution matrix obtained from transmit pulse samples \textbf{s} defined by
\[
\textbf{S} = \begin{bmatrix}
s_{0}       & 0 & \hdots & \hdots & 0 \\
\vdots      & s_{0} &  &  & \vdots \\
s_{N-1}     & \vdots & \ddots & & \vdots \\
0       & s_{N-1} &\ddots  &  & \vdots \\
\vdots & & \ddots & \ddots & 0\\
 & & & & &\\
\vdots & & & \ddots & \vdots\\
0 & \hdots & & \hdots & s_{N-1}\\

\end{bmatrix}
\]
The size of \textbf{S} is $(L+N-1)XL$ where L is the number of columns of the filter weight vector, \textbf{W}, and N is the number of samples in transmit waveform. We form the desired output vector, \textbf{d} and compare this with the actual output \textbf{y} of the filter. The filter output is given by
\begin{equation}
\textbf{y} = \textbf{S}\textbf{W}
\end{equation}
The LS formulation of the adaptive filter would minimize the mean square error between the desired output \textbf{d} and the actual output \textbf{y} as the cost function and yield the optimum filter weights as
\begin{equation}
\textbf{W}_{LS} = (\textbf{S}^{H}\textbf{S})^{-1}\textbf{S}^{H}\textbf{d}
\end{equation}

This can be recursively achieved through Recursive least squares (RLS) algorithm. So we set up the desired response \textbf{d} and the RLS iteratively tries to minimize the MSE.\par

The use of minimum integrated sidelobe (ISL) filter already shows promising result towards achieving this goal. The ISL filter is the one that minimizes the sidelobe energy by constructing a modified transmit convolution matrix by deleting the columns of \textbf{S} that corresponds to the mainlobe of the ambiguity function obtained from \textbf{y} \cite{Bharadwaj2012}. This yields the optimum filter weights as

\begin{equation}\label{isl}
\textbf{W}_{ISL} = \dfrac{\alpha(\textbf{S}_{m}^{*}\textbf{S}_{m}^{T})^{-1}\textbf{s}^{H}}{\textbf{s}(\textbf{S}_{m}^{*}\textbf{S}_{m}^{T})^{-1}\textbf{s}^{H}}
\end{equation}
where $\textbf{S}_{m}$ is the modified transmit convolution matrix and $\alpha$ is a constraint on the peak mainlobe power.

\begin{figure}[!t]
	\centering
	\includegraphics[width=2.5in]{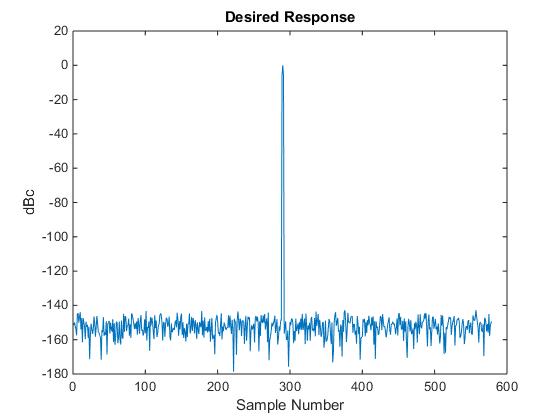}
	\caption{Desired Response for the RLS algorithm}
	\label{fig_sim1}
\end{figure}

\begin{figure}[!t]
\pagestyle{empty}

\tikzstyle{decision} = [ diamond, aspect=1, draw, fill=blue!20, text width=4em, text badly centered, node distance=2.5cm, inner sep=-5pt ]
\tikzstyle{block} = [ rectangle, draw, fill=blue!20, text width=20em, text centered, rounded corners, minimum height=4em ]
\tikzstyle{block1} = [ rectangle, draw, fill=blue!20, text width=4em, text centered, rounded corners, minimum height=4em ]
\tikzstyle{line} = [ draw, -latex' ]

\begin{tikzpicture}[node distance=2cm, auto]\label{flowchart}

\node [block] (init) {Initialise BW, Duration of pulse, sampling rate} ;
\node [block, below of=init] (block-1) {Generate the 480 tap min ISL filter} ;
\node [block, below of=block-1] (block-2) {Define the desired response for RLS algorithm} ;
\node [block, below of=block-2] (block-3) {Initialise the parameters of the RLS algorithm and the weight vector to be the min ISL weight vector} ;
\node [block, below of=block-3] (block-4) {Start the iterations} ;
\node [block, below of=block-4] (block-5) {Calculate the ISL with updated weight vector at every iteration step} ;
\node [decision, below of=block-5] (decision-1) {Is the ISL minimum?} ;
\node [block1, right of=decision-1] (block-6) {Save the iteration number and the weight vector} ;
\node [block1, left of=decision-1] (block-7) {Continue Iterations} ;

\path [line] (init) -- (block-1) ;
\path [line] (block-1) -- (block-2) ;
\path [line] (block-2) -- (block-3) ;
\path [line] (block-3) -- (block-4) ;
\path [line] (block-4) -- (block-5) ;
\path [line] (block-5) -- (decision-1) ;
\path [line] (decision-1) -- node {yes}(block-6) ;
\path [line] (decision-1) -- node {no}(block-7) ;

\end{tikzpicture}
\caption{The flow chart representation for an RLS optimizer for ISL.}
\label{fig_sim2}
\end{figure}
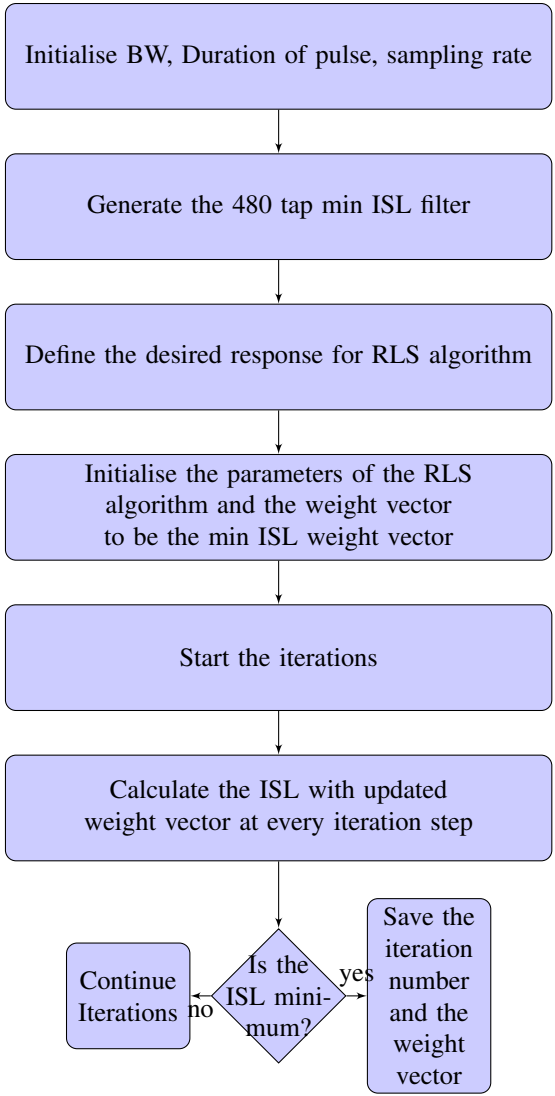

The training part of the RLS adaptive filter involves presenting the rows of the convolution matrix \textbf{S} along with the output from the desired vector \textbf{d}. This is done iteratively till the wieght converges to minimum mean square error estimate (MMSE) between desired and the actual output \textbf{y}. The ISL is defined by:
 
\begin{equation}
ISL = \textbf{W}\textbf{S}_{m}^{*}\textbf{S}_{m}\textbf{W}
\end{equation}

and this gives the performance metric for achieving optimal weights. The lesser is the ISL, the better sidelobe suppression is being achieved. So at each iteration, we monitor the ISL. When we plot ISL versus iteration number, we can easily identify the iteration number where the ISL was minimum. This would correspond to the MMSE solution of RLS adaptive algorithm. The desired response has been set as shown in Fig. \ref{fig_sim1}. It is easy to observe that the desired solution has no sidelobes and the mainlobe consists of three samples. The initial weight vector is set to the minimum ISL solution, $\textbf{W}_{ISL}$ of equation \eqref{isl}. The full method is highlighted in the flowchart depicted in Fig. \ref{fig_sim2}.

\subsection{Results}
\begin{figure}[!t]
	\centering
	\includegraphics[width=3in]{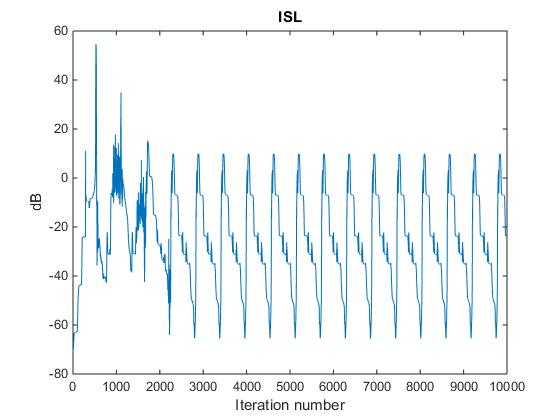}
	\caption{Integrated sidelobe level variation with iterations}
	\label{fig_sim3}
\end{figure}

The initial weight vector for the RLS optimization algorithm highlighted in the flowchart above was obtained with the following parameters in mind: BW = 5MHz, pulse width = $20 \mu s$, mainlobe width = 3 samples, tuckey window parameter $\alpha = 0.1$ and the filter length = 480 coefficients. We ran RLS iterations for about 10,000 times and the computed ISL at each iteration which is plotted in Fig. \ref{fig_sim3}.

\begin{figure}[!t]
	\centering
	\includegraphics[width=3in]{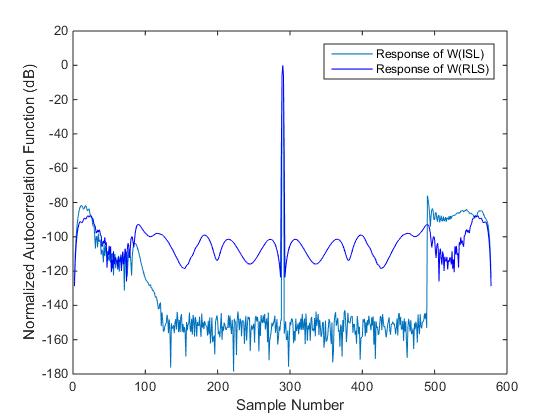}
	\caption{Comparision of sidelobe levels after convolving with Weights obtained from min ISL and weights obtained after RLS optimization.}
	\label{fig_sim4}
\end{figure}

We see that after approximately iteration number close to 2200, the ISL is varying periodically. We obtain the iteration number where the ISL was minimum and use that weight vector for the pulse compression filter (which is a mismatched filter based). The FIR filter response to the chirp signal for the weight vector where ISL was found to be minimum is shown in Fig.\ref{fig_sim4}, additionally with the response from $\textbf{W}_{ISL}$, that we originally started with.

It can be clearly observed from this figure that the filter weights obtained after RLS and convolved with chirp, has very low level of sidelobes in the region around the mainlobe. So there is a lot of sidelobe free region. This performance is comparable and even better than the cyclic algorithms for polyphase codes proposed in \cite{Haohe2009} and \cite{Deng2004}. 

\subsection{Use of CLEAN deconvolution to get rid of high power echoes}
The timing portion in NASA D3R radar, which is a dual-frequency, dual polarization, Doppler radar has high power transmitter injection as an online means for transmit calibration. This serves to track the variation of transmit power on a pulse by pulse basis. If such a high power transmit pulse is convolved with large lengths pulse compression filters (mismatched), the sidelobes may have impact on the received echo samples which are even at a far range from the radar. Additionally, we can have a similar effect with reflections from near by buildings and ground clutter, which will be high power echoes in near range but they have the potential to spill out sidelobe power to far range, which means that their effect can be seen even at far distance due to sidelobe power from long length mismatched pulse compression filters. We need to deconvolve such high SNR targets from our reflectivity profiles, so that the sensitivity could be enhanced.\par
D3R has undergone extensive field trials in the past. Refer to \cite{Chandrasekar2016}, \cite{Chandrasekar2017a}, \cite{Chandrasekar2017b}, \cite{Chandrasekar2018} and \cite{Chandrasekar2019} for performance of D3R in various campaigns. Recently D3R has gone through extensive upgrade in hardware of digital receiver, IF up-converters ad down-converters to allow larger filter lengths and flexible waveforms (\cite{kumar2017}, \cite{kumar2018} and \cite{kumar2020}).

\subsubsection{CLEAN Detector}
We begin with the assumption that all high power echoes whose sidelobes are capable to have an adverse impact on the far targets with smaller SNR, have been detected. This can be accomplished through a much smaller length matched filter. If the actual impulse response of a scatterer is $\textbf{a}$, then the output of the pulse compression filter would be 

\begin{equation}
\textbf{y} = \textbf{S}\textbf{a} + \textbf{n}
\end{equation}
where \textbf{n} is the receiver noise vector (matched filter assumption). The clean deconvolver, is the LS estimator for this scattered impulse response \textbf{a} and an efficient estimator is given by:
\begin{equation}
\textbf{a} = (\textbf{S}\textbf{S}^{H})^{-1}\textbf{S}\textbf{y}
\end{equation}

This represents the deconvolution of \textbf{y} to obtain an efficient estimate of \textbf{a}.

In \cite{Foreman2010}, the clean detector has been derived as summation of the signal power due to a scatterer at that range cell and the interference power due to scatterers at other range cells. He formed a hypothesis $\textbf{H}_{0}$ where only interference is present as:
\begin{equation}
\textbf{y}_{I} = \textbf{S}\textbf{a} + \textbf{n}
\end{equation}
and alternate hypothesis $\textbf{H}_{1}$ that signal plus interference due to other scatterers is present at the $k^{th}$ range cell given by:

\begin{equation}
\textbf{y} = \textbf{S}\textbf{$\delta$}_{k} A + \textbf{y}_{I}
\end{equation}
where $\textbf{$\delta$}_{k}$ is the column vector denoting the test for the  $k^{th}$ scatterer and A is its amplitude.He proposed the following test for presence of only interference (or signal plus interference) in case the target returns are not correlated:
\begin{equation}
\left| \textbf{$\delta$}_{k}^{H}\textbf{S}^{H}(\textbf{S}\textbf{S}^{H} + \sigma^{2}_{n}\textbf{I})^{-1}\textbf{y}\right| \lessgtr \eta
\end{equation}

where $\eta$ is the probability of false alarm. It can be observed that the operation, $\textbf{$\delta$}_{k}^{H}\textbf{S}^{H}(\textbf{S}\textbf{S}^{H})^{-1}$ on $\textbf{y}$ is a deconvolution operation (with a matched filter). If $\textbf{b}$ is a vector that contains the large magnitude scatterers whose sidelobes need to be deconvolved, then the best estimate of the impulse response \textbf{a} is

\begin{equation}
\textbf{\^{a}} = \textbf{$\delta$}_{k}^{H}\textbf{S}^{H}(\textbf{S}\textbf{b}^{H}\textbf{b}\textbf{S}^{H} + \sigma^{2}_{n}\textbf{I})^{-1}\textbf{y}
\end{equation}

This provides a good estimate of target amplitude in the range cell under consideration, by eliminating the sidelobes of larger targets (in other range cells) under low SNR condition. This technique was originally used to clean up optical images where the target of interest was blurred by the point spread function (PSF) of the instrument. Next it was gradually applied for microwave imaging and now in radar technology as well. Thus the CLEAN deconvolution can also be stated as an estimation problem where we try to estimate amplitude and phase of all detected targets accurately in a sidelobe free environment.

\section{Non-Linear Frequency Modulation(NLFM) Compression Waveform for Improved Sensitivity}
Till now, all the results were presented with LFM (chirp based) waveforms. However, NLFM based waveform design is not restricted to a fixed sweep rate as in Linear frequency modulation. The frequency function is linear in case of LFM and because of the non-linear frequency function, there is more flexibility in case of NLFM waveform with the tapering incorporated in frequency rather then in amplitude to achieve sidelobe reduction. In case of LFM, windowing leads to significant degradation of SNR and sensitivity loss if severe window functions are applied. The normalized two-way SNR loss can be characterized by: 

\begin{equation}
SNR_{loss} = 10log_{10}(\dfrac{(\sum_{n=1}^{N}w_{t}w_{r})^{2}}{N[\sum_{n=1}^{N}(w_{t}w_{r})^{2}]})
\end{equation}

where $w_{t}$ and  $w_{r}$ is the amplitude weighting function for the transmit and receive section respectively and N is the length of the pulse in samples. \par
In case of NLFM waveform, the frequency function can be shaped by having a sharper roll-off of frequency along the edges of the frequency function. However, still this would lack flexibility for a global optimization technique such as a genetic algorithm. This goal can be fulfilled by the use of Bezier curves in the frequency function of the NLFM waveform design.

\subsection{Bezier Curves}
Bezier Curves are parametric functions and characterized by using the same base function for all its dimensions. The base function is represented by:

\begin{equation}
Bezier(n,t) = \sum_{i=0}^{n}\binom{n}{i}(1-t)^{n-i}t^{i}
\end{equation}
where n is the order of the Bezier curve and t is the time index. If we want to change the curve, we need to change the weight of each point, effectively changing the interpolations, as illustrated below:
\begin{equation}
Bezier(n,t) = \sum_{i=0}^{n}\binom{n}{i}(1-t)^{n-i}t^{i}.w_{i}
\end{equation}
These weights act as the control points that define the curvature of the Bezier curve. These control points can be used to give deviation from the regular linear frequency function of the LFM pulse to gain better characteristics of NLFM with the desired features (lower sidelobe etc).
Sometimes, the Bezier base function coefficients can be stored and read from memory, instead of computing them, making it computationally simple.
We utilize ten control points along the line spanning half the pulse width. Another fixed two points are added as the start and end points for the curve. The other half of the frequency function is mirror image of the first half, so as to have the Doppler tolerance capability in the NLFM. Because of the large search space, genetic algorithms (GA) were selected.

\subsection{The use of Breeder genetic algorithms(BGA) to find optimum frequency function}
GA has been the most popular technique in evolutionary computation research. Holland proposed GA as a heuristic method based on "survival of the fittest" and since then GA has been proved to be a useful tool for search and optimization. In GA terminology, the space of all feasible solutions is called search space or state space. Each possible solution has a fitness value based on the problem definition. With GA, one looks for the best solution among number of different solutions, based on fitness values, represented by one point in the search space (Intro to GA, 2008).\par
The population members are randomly paired to create progeny (also known as Crossover). An important feature of GA is the ability to avoid local maxima in fitness function by inducing mutation in population. The use of mutation results in a higher likelihood that the search will not stall at local maxima.\par
Next we need to define a fitness function which is the goal of optimization of GA. The Integrated sidelobe level defined by equation 4, is selected as the fitness function to be evaluated for each population member and at every iteration. The GA tries to minimize this function. The GA process repeats until the stopping criterion is met. Typically, the stopping criterion is a minimal change in average fitness over successive iterations or a number of generations without improving the fitness function.\par
Forcing symmetry in the frequency function of the NLFM would lead to better Doppler tolerance. So the degree of freedom to the GA are the ten control points that define the curvature of the Bezier curve in half of the frequency function space of the NLFM. The rest two control points are the start and end of the curve and are thus fixed points. The rest half is just the mirror image. The GA tries to optimize these ten control points of the Bezier curve, forming the non-linear frequency function and evaluates fitness score at each iteration for each of the individual. The fitness is the ISL function for the mismatched filter implementation. In this way, GA optimizes the frequency function in such a way so as to achieve better ISL at each generation by utilizing mutation and crossover of the best individuals in a generation pool.\par
We plan to use breeder genetic algorithm(BGA). It is based on artificial selection similar to that used by human breeders. Selection and mutation are analyzed within this framework. It is a well known optimization technique. The steps involved in the BGA process for our optimization formulation are given below:\\
1) Define the genetic representation of the problem. It is the ten control points of the Bezier Curve in our case.\\
2) Create an initial population $P(0)$ of size N. \\
3) Pair the pairs at random forming N pairs.\\
4) For each pair, we apply crossover operator to produce two offspring followed by the mutation operator. This yields the next generation $P(t+1)$. \\

The genetic structure of the problem consists of ten control points of the Bezier curve that gives curvature to the frequency function of the NLFM. The control points are able to take in continuous values from 0 to Bandwidth/2. Thus the initial population consists of vectors of length ten containing random numbers between 0 and Bandwidth/2. N such vectors are constructed. The larger is the population size, the better chances for reaching the solution quickly. Lets discuss the mutation operator for our problem set where in the genes can take in continuous values.

\subsubsection{Simulations and Results with BGA}
\begin{figure}[!t]
	\centering
	\includegraphics[width=3.5in]{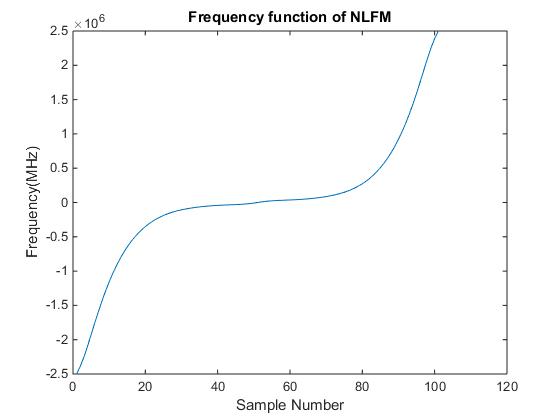}
	\caption{Frequency Function}
	\label{fig_sim5}
\end{figure}

\begin{figure}[!t]
	\centering
	\includegraphics[width=3.5in]{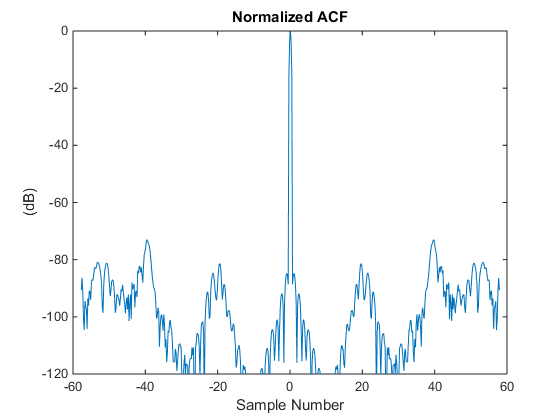}
	\caption{ACF}
	\label{fig_sim6}
\end{figure}

\begin{figure}[!t]
	\centering
	\includegraphics[width=3.5in]{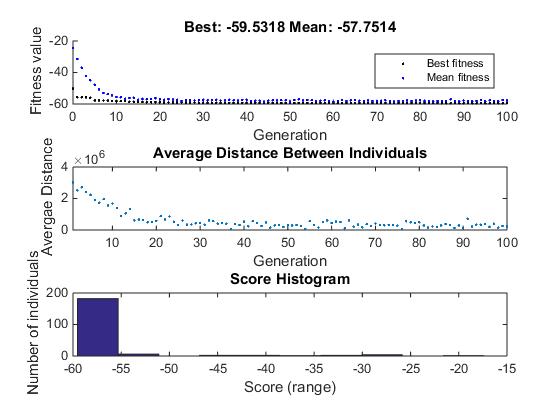}
	\caption{Summary of the BGA Optimization}
	\label{fig_sim7}
\end{figure}

The parameters taken for simulation are: Pulse Width ($\tau$) = 20$\mu$s, Bandwidth = 5MHz and ten control points for the Bezier curve. For each individual in the population at a given generation, the bezier curve is evaluated and the frequency function becomes known. Then the NLFM waveform is formed. After this, the mismatched filter is evaluated for this waveform and the ISL is computed. The ISL is the metric which tells us about the performance of the NLFM waveform and is to be minimized through the BGA. The BGA uses population size of 200 and top 40\% of the parents on the fitness scale are selected for mating and crossover. We consider only single point crossovers. The mutation rate is set to uniform 0.1\% so that the algorithm can escape local minima.
The Figure below shows the best fitness and the average fitness progression as the optimization progress. The average euclidean distance between population decreases as more and more fit individuals are allowed to reproduce to form the newer generation. We also show the best individual frequency function and the auto-correlation function of the waveform with the mismatched filter generated after the optimization in Fig. \ref{fig_sim5}, \ref{fig_sim6} and \ref{fig_sim7}.


If we try to compare chirp based adaptive filter scheme with NLFM based one, we can observe that amplitude tapering in chirp based filters can lead to SNR loss which doesn't happen for NLFM based mismatched filters, however the peak sidelobe levels achieved with chirp based filters is much better. We have not tried but a joint optimization of Bezier curve of NLFM and adaptive filters for getting mismatched coefficients will likely give better performance of sidelobe levels and SNR loss. This is a good research area for further studies.

\section{Conclusion}
The two techniques for sidelobe level reduction with adaptive filter approach were discussed and their relative pros and cons were discussed. Finally it looks like a joint optimization of Bezier curve of NLFM and adaptive filters for getting mismatched coefficients will likely give better performance of sidelobe levels and SNR loss.

\ifCLASSOPTIONcaptionsoff
  \newpage
\fi

\end{document}